\documentclass[11pt,twocolumn,twocolappendix]{aastex631}
\usepackage{comment}
\usepackage{amsmath}
\usepackage{wrapfig}
\usepackage{mathtools}
\graphicspath{{figs/}}
\usepackage{hyperref}
\usepackage{nameref}
\usepackage{float}
\usepackage{graphicx}
\usepackage{natbib}
\begin{document}

\title{Evolution of Stars During the Main Sequence and the Transition to the Red Giant Phase}

\shorttitle{Red Giants}
\author{Ravid Achituv \& Re'em Sari}
\affiliation{Racah Institute of Physics, The Hebrew University of Jerusalem, Jerusalem 91904, Israel}

\email{Email: ravid.achituv@mail.huji.ac.il}

\begin{abstract}
We derive a simple analytical description for the structure and evolution of $3$--$10\,M_\odot$ stars throughout main-sequence hydrogen burning. We obtain an analytical relation for the convective core mass,\(
\frac{M_{\star}}{M_c} = 1+2.1(\frac{\mu_c}{\mu_e})^2,\)
where $\mu$ is the mean molecular weight of the core and the envelope. Using this relation, we analytically derive the hydrogen abundance profile outside the convective core.
We find that $\mu(m) \propto m^{-0.7}$, and show that this profile is important for an analytical description of these stars. Within this region of variable $\mu$, the temperature, density and pressure are well approximated by power laws of the radius.
We derive analytical expressions for the core and stellar radii, stellar luminosity, and effective temperature as functions of $\mu_{\rm c}$.

We provide a simple physical explanation for the main-sequence ``hook'', defined by the minimum in the effective temperature. 
We show that the hook occurs when the mass fraction of hydrogen in the core is $x_{\rm c}\simeq0.045$, and stress that the same convective-core burning physics govern the subsequent evolution. In that sense, at the ``hook" hydrogen is not yet fully exhausted. 
During late main-sequence evolution we find that the ratio of nuclear luminosity between the core and the surrounding hydrogen-rich shell is $\simeq4000x_c$, hence the main sequence terminates only once $x_{\rm c}\simeq2.5\times10^{-4}$, when the surrounding layers become as luminous as the core itself and $M_{\rm c}\simeq0.11\,M_\star$. 

Although this terminal core mass is numerically similar to the Sch\"onberg--Chandrasekhar limit, we show that the two are physically unrelated, since the core remains far from isothermal even at this stage. We validate all analytical results using MESA simulations.

\end{abstract}
\section{Introduction}

Throughout their main-sequence stage, stars evolve as hydrogen in their cores burns into helium and the mean molecular weight of the core increases. 
Stars more massive than $\sim2\,M_\odot$ follow similar tracks in the H--R diagram  \citep{Kippenhahn2012}. Their luminosities and radii can change by as much as a factor of two, and their H--R diagram track includes a minimum in the effective temperature, known as the main-sequence ``hook" \citep{Pols1998,Sabhahit2025}. 
In these stars, the dominant hydrogen-burning process is the temperature-sensitive CNO cycle. Therefore, the energy generation is concentrated in a small region at the center of the star. Yet, the burned elements are mixed by convection into a chemically homogeneous core.

\begin{figure}[htbp]
\includegraphics[
    width=1.05\linewidth,
    trim=1.5cm 0cm 0cm 0cm,
    clip
]{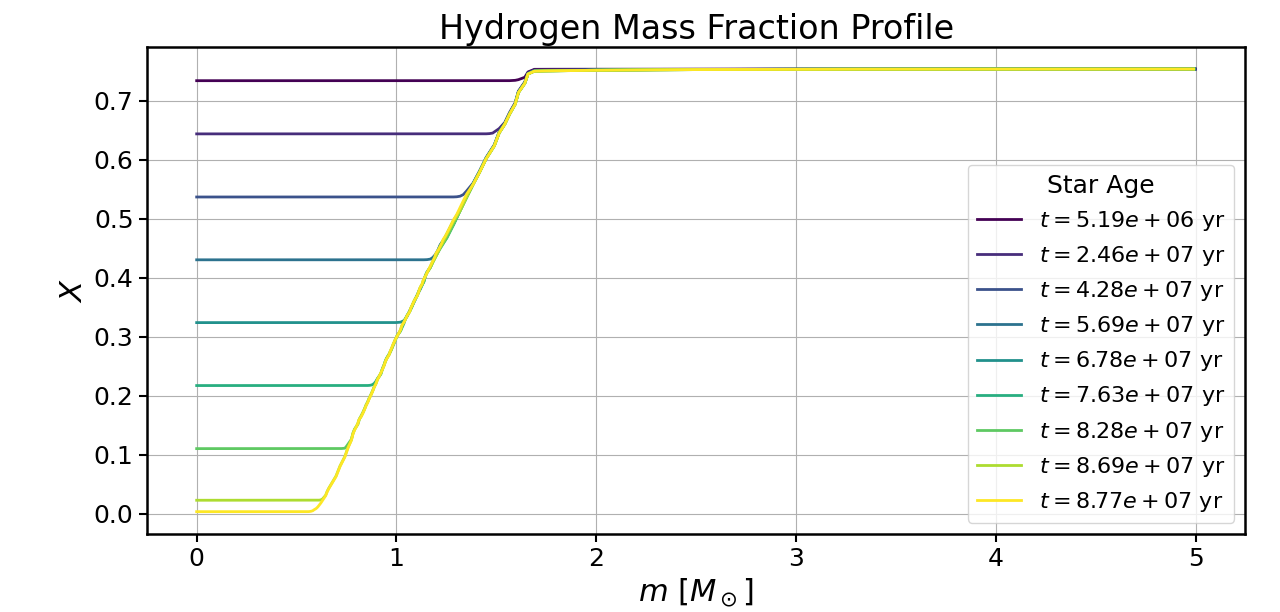}
   \caption{Hydrogen mass fraction $x$ as a function of the mass coordinate $m$ for a $5\,M_\odot$ main-sequence star at different times. 
The straight lines at left represent the convective core, which is fully mixed. 
The $x(m)$ profile between the core and the envelope develops because of the convective core evolution.}

    \label{fig:mu}
\end{figure}

The convective core decreases in both mass and radius during main-sequence evolution.
As the core shrinks in mass, it leaves behind a distinct chemical-composition profile  \citep{Kippenhahn2012}, as shown in Fig.~\ref{fig:mu}.

Although the shrinking of the convective core has been well studied numerically, we aim to derive analytical expressions that describe it. 
This analytical description provides a good estimate of the helium-core mass at the end of the MS. This is useful, for example, to estimate the remnant white dwarf mass formed in close binary systems \citep{Tauris2000,Tauris2011,ShaoLi2012,Misra2020,BareliGinzburg2026}.

During late stages of the main sequence, when the hydrogen abundance in the core, $x_{c}$, becomes very low, a burning shell begins to develop outside the core. The temperature in this shell is lower than in the core, but the hydrogen abundance is much higher. Our formalism allows us to determine the value of $x_{\rm c,end}$ at which the shell produces the same amount of luminosity as the core. We take this point to mark the end of the main sequence. There is no complete core hydrogen exhaustion, in the sense that the core does not stop burning before the surrounding shell ignites. 


In this work we focus on stars in the mass range of $3$--$10\,M_\odot$. The lower limit of $3\,M_\odot$ is chosen so that the energy production is dominated by the CNO cycle, which we approximate \citep{iliadis2007nuclear,kippenhahn2012stellar} as
\begin{equation}
\epsilon \left[\frac{\rm erg}{\rm s\,g}\right]
\simeq \epsilon_1 x z \rho T^\nu, 
\label{eq:epsilon}
\end{equation}
where $z$ is the metallicity, $x$ is the hydrogen mass fraction, and we adopt
$\nu=14,\epsilon_1=1.05\times10^{-99}[\frac{erg\,cm^3}{s\,g^{2}\,K^{14}}]$. These values agree within 10\% with those used in our MESA simulations.
The upper limit of $10\,M_\odot$ is chosen because, within this mass range, the pressure is dominated by gas pressure
 \begin{equation}
     P=\frac{k_B\rho T}{\mu m_H},
     \label{eq:pgas}
 \end{equation}
where $k_B$ is the Boltzmann constant, $m_H$ is the proton mass and \(
\mu
= \frac{4}{5x+3}
\) is the dimensionless mean molecular weight. 
Stars with larger masses are dominated by radiation pressure, which requires a different treatment.

In the mass range considered here, Thomson opacity, $\kappa \cong 0.2(1+x)\ {\rm cm^2\,g^{-1}} $, dominates throughout most of the star.

To validate and guide our results we use the stellar evolution code \textsc{MESA} \citep{Paxton2011,Paxton2013,Paxton2015,Paxton2018,Paxton2019,Jermyn2023}, version r24.08.1. We use the \texttt{7M\_prems\_to\_AGB} test suite to construct stellar models with masses in the range $3$--$10\,M_\odot$. 
We use an exponential overshoot prescription \citep{herwig2000,paxton2013mesa} with $f=0.014$ and $f_0=0.004$. We adopt a metallicity of $Z=0.002$, and compute the nuclear energy generation self-consistently in \textsc{MESA} using the \texttt{o18\_and\_ne22.net} reaction network.

The structure of this paper is as follows: In Section~\ref{sec:mcore}, we present a simple two-layer model of MS stars. In Section~\ref{sec:3layers}, we use the simple model to construct a more accurate model consisting of an additional intermediate layer where the chemical composition smoothly changes between that spatially constant value of the core to that of the unburned envelope. In Section~\ref{sec:shell} we discuss the end of the MS and find the terminal value of $x_{\mathrm{c,end}}$. In Section~\ref{sec:dis} we discuss the limits of our model, and in Section~\ref{sec:sum} we summarize our results.

\section{Two-layer model}
\label{sec:mcore}
For a simple description of the star, we use a model that consists of two regions: (i) a convective core with $\mu_c,M_c,R_c$ and central conditions $T_c,\rho_c$; and (ii) a radiative envelope with $\mu_e,M_e,R_e$, temperature and density at the beginning of the envelope $T_e,\rho_e$, and constant luminosity $L_{\star}$. This model neglects the intermediate region between the core and the envelope in which a $\mu(m)$ profile develops. We treat this region more carefully in section \ref{sec:3layers}.
Ignoring overshooting, the transition condition between the two regions is that the nuclear luminosity generated in the core can be transported by radiation
\begin{equation}
\begin{split}
    L_{\mathrm{nuc}}
    &= \int_0^r \epsilon \, 4\pi \rho r^2 dr
    = r^2 \frac{ac}{3\kappa\rho} \frac{dT^4}{dr}
    = L_{rad}.
\end{split}
\label{eq:ths}
\end{equation}
For each region, we use
\begin{equation}
    M_c\simeq \rho_cr_c^3,   M_e\simeq \rho_er_e^3,
    \label{eq:1}
\end{equation}
and from hydrostatic equilibrium
\begin{equation}
    r_{c} \simeq \sqrt{\frac{k_B}{ G \mu_{c}m_p \rho_c}T_c},  r_{e} \simeq \sqrt{\frac{k_B}{ G \mu_{e}m_p \rho_e}T_e} .
        \label{eq:2}
\end{equation}
We require continuity of pressure and temperature between the edge of the core ($\rho_{ce}\simeq \rho_c,  T_{ce}\simeq T_c$) and the beginning of the envelope, such that
\begin{equation}
\frac{\rho_e}{\rho_c}\simeq\frac{\mu_{\rm e}}{\mu_{\rm c}}\qquad T_e\simeq T_c.
    \label{eq:3}
\end{equation}
In Appendix A, we solve this set of equations \eqref{eq:ths}-- \eqref{eq:3} while including factors of order unity. We show that these factors are approximately constant during the main sequence. 

Using equations \eqref{eq:1}, \eqref{eq:2}, and \eqref{eq:3}, we obtain the ratio of the masses
\begin{equation}
\frac{M_{e}}{M_{c}}
=
\frac{M_{e,i}}{M_{c,i}}\left(\frac{\mu_{\rm c}}{\mu_{\rm e}}\right)^{2}\simeq 2.1\left(\frac{\mu_{\rm c}}{\mu_{\rm e}}\right)^{2},
\label{eq:MenvMc_simple}
\end{equation}
where \(\frac{M_{e,i}}{M_{c,i}}\) is the ratio between the envelope and core masses at ZAMS. In our MESA simulations, for the range $3$--$10\,M_\odot$, we find \(1.6<\frac{M_{e,i}}{M_{c,i}}<2.5\), and we adopt $\frac{M_{e,i}}{M_{c,i}}=2.1$. 
The small variation in this value over the stellar-mass range of interest is mainly due to differences in the importance of radiation pressure at the core, which we ignore.

Finally, requiring the total stellar mass to remain constant gives for the core mass
\begin{equation}
M_{c}=\frac{M_{\star}}{\left(1+2.1\left(\frac{\mu_{c}}{\mu_{e}}\right)^{2}\right)}.
\label{eq:Mc_from_Mtot_simple}
\end{equation}
As a first approximation to the core-mass evolution, we take for the envelope the ZAMS value $\mu_e=0.6$. For a more accurate approximation, we use this first approximation namely the $\mu(m)$ profile implied by equation~\eqref{eq:Mc_from_Mtot_simple} with $\mu_e=0.6$, to compute a mass-weighted average value, $\langle\mu\rangle_e$, for the envelope. This average includes both the region outside the core between the initial and the present core masses, and the outer envelope. The detailed calculation is given in Appendix B. In Fig.~\ref{fig:Mcore}, we plot equation~\eqref{eq:Mc_from_Mtot_simple} with these two choices of $\mu_e$ and compare them with the $M=5\,M_\odot$ MESA simulation. 
Our formula yields good results, where at the end of the MS the error between the simulation and equation~\eqref{eq:Mc_from_Mtot_simple} with average $\langle\mu\rangle_e$ is $\sim 10\%$.  The main reason for this deviation is that in order to derive equation~\eqref{eq:Mc_from_Mtot_simple}, we use equations \eqref{eq:1}  and \eqref{eq:2} that assume that the core mass is small enough compared to the envelope. This assumption is not accurate enough, especially at the beginning of the MS, where $M_c \cong M_e/2.1$.
\begin{figure}[htbp]
\includegraphics[
    width=1.05\linewidth,
    trim=2.5cm 0cm 0cm 0cm,
    clip
]{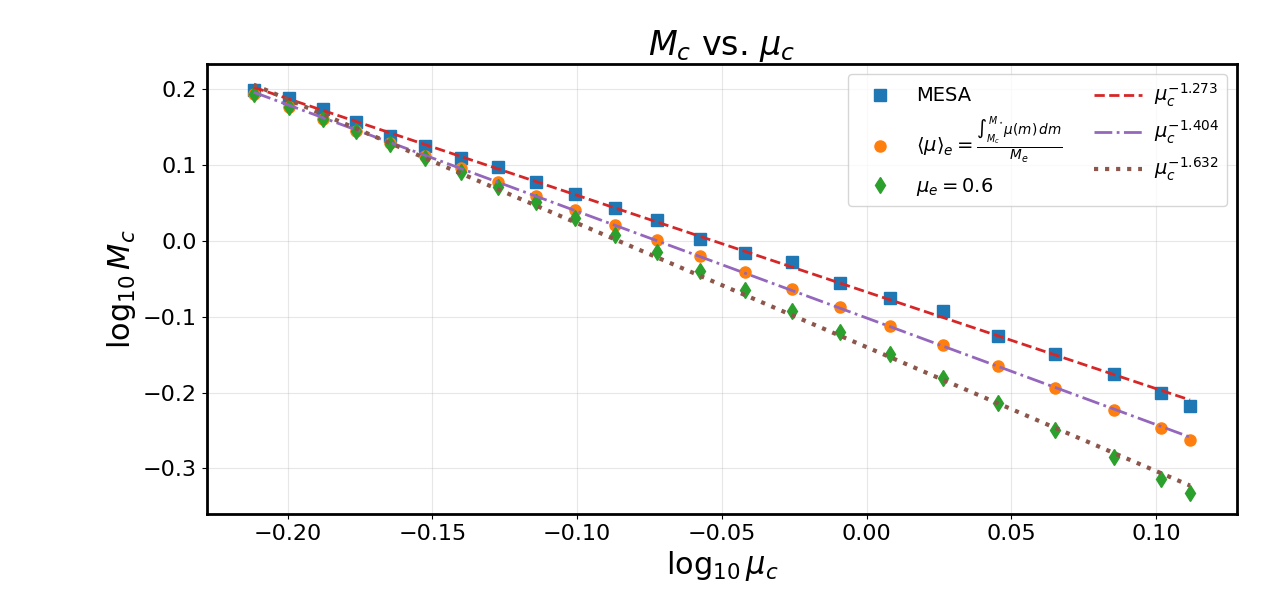}
    \caption{The convective core mass, as a function of the central $\mu_c$ in log scale.
    In blue we show the data from our MESA run for $M=5\,M_\odot$. In green and orange we show equation \eqref{eq:Mc_from_Mtot_simple} with constant $\mu_{e}=0.6$ and averaged value $\langle\mu\rangle_{e}$, respectively. For simplicity, we fit each curve with a power law.
    }
    \label{fig:Mcore}
\end{figure}

 At the end of the MS, $\mu_c$ is significantly higher than $\mu_e$ and equation \eqref{eq:Mc_from_Mtot_simple} can be approximated by 
 \begin{equation}
 M_c=0.47(\frac{\mu_e}{\mu_c})^2M_{\star}\simeq 0.11M_{\star}.
 \label{eq:Mc_final_frac}
 \end{equation}
In the last step, we substituted $\mu_c=4/3$ to obtain the terminal core mass. 
 This value is larger than the Schönberg--Chandrasekhar mass limit, \(M_{sc}=0.37(\frac{\mu_e}{\mu_c})^2M_{\star}\). Schönberg and Chandrasekhar \citep{Schonberg1942} used a similar model to the model we present, but an isothermal core rather than a convective core. This difference leads to a smaller coefficient for their core mass. In Section \ref{sec:shell}, we show that an isothermal core is not relevant for describing the end of the MS for stars in the mass range considered here, contrary to what was suggested by some authors \citep{Kippenhahn2012,Ziolkowski2020}.

For the remainder of this paper, in order to analytically follow the various aspects of the stellar evolution, we use a power-law fit for the core mass as a function of $\mu_c$. A fit to equation \eqref{eq:Mc_from_Mtot_simple} with a constant $\mu_e=0.6$  yields an exponent of $-1.63$ , while using the average envelope mean molecular weight $\langle\mu\rangle_e$ yields an exponent of  $-1.40$. Instead, we choose to use our MESA simulation and find in the MS $\mu_c$ range $0.6<\mu_c<4/3$ 
\begin{equation}
    {M_c}\propto {\mu_c}^{-1.27},
\label{eq:mfit}
\end{equation}
as shown in Fig.~\ref{fig:Mcore}.

In this $\mu$ range we also fit a power law for the opacity used in our MESA simulations, which is dominated by Thomson opacity ($\sim 93\%$), but also includes absorption for partially ionized heavy elements such as iron, and therefore also depends slightly on the temperature
\begin{equation}
\kappa\propto\mu^{-0.82}T^{0.08},
    \label{eq:kfit}
\end{equation}
as shown in Fig.~\ref{fig:kappa}.

\begin{figure}[htbp]
\includegraphics[
    width=\linewidth,
    trim=2.5cm 0cm 0cm 0cm,
    clip
]{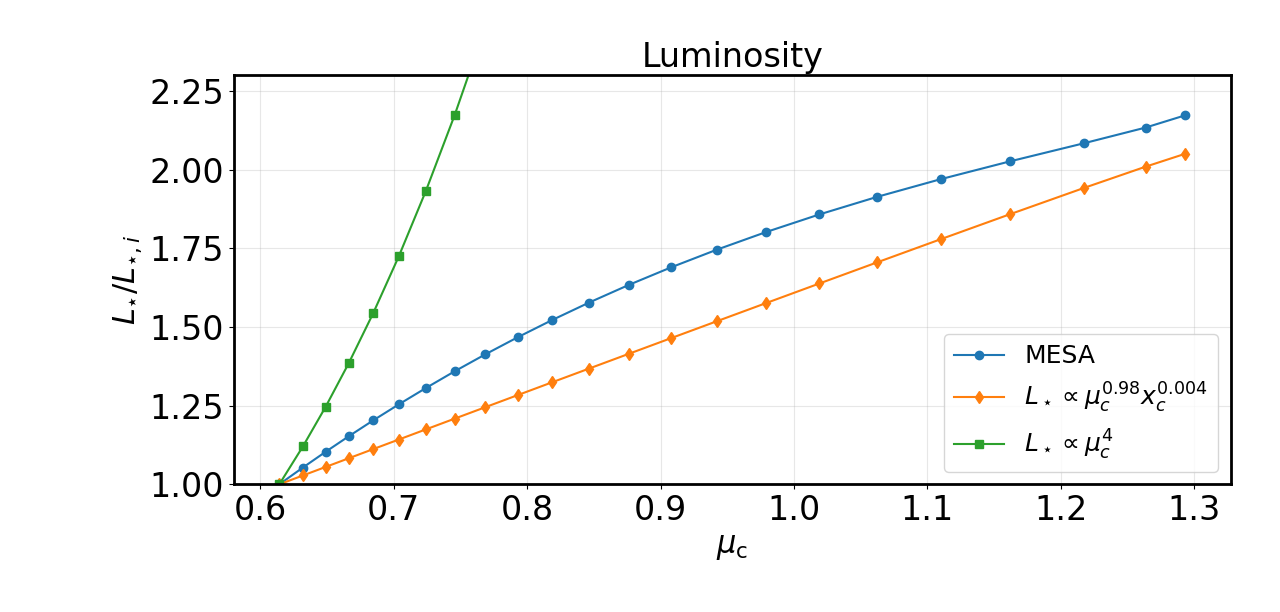}
    \caption{Stellar luminosity divided by its ZAMS value as a function of the central $\mu_c$. We compare equation \eqref{eq:Lfinal} in orange, with the MESA simulation for $M=5M_{\odot}$. For comparison, we show in green the luminosity derived for a chemically homogeneous star, $L_{\star}\propto\mu_c^4$ \citep{kippenhahn2012stellar}, which is not an accurate description of intermediate and massive MS stars.}
    \label{fig:LLL}
\end{figure}

Using equations \eqref{eq:ths}, \eqref{eq:2}, and \eqref{eq:mfit}, we obtain the central temperature
\begin{equation}
T_c\propto \mu_c^{14/85}x_c^{-1/17},
\label{eq:ttt}
\end{equation} 
and the core radius
\begin{equation}
    R_c\propto 
    \mu_c^{-17/40}x_c^{1/17},
    \label{eq:rrr}
\end{equation}
as shown in detail in Appendix A.
Using these power laws \eqref{eq:mfit}--\eqref{eq:ttt}, and equations (\ref{eq:ths})--(\ref{eq:2}) at the edge of the core, we derive for the luminosity:
\begin{equation}
    L_{\star}
    \propto\frac{{M_{c}^3 \mu_{c}^4}}{\kappa}\propto\mu_c^{0.98}x_c^{0.004}\propto\mu_c.
    \label{eq:Lfinal}
\end{equation}
 We compare this result with our MESA simulation in Fig.~\ref{fig:LLL}.
 
 Using equations~\eqref{eq:Lfinal} and \eqref{eq:mfit}, we obtain the evolutionary timescale at any stage along the main sequence:
\begin{equation}
\begin{aligned}
t(x)\equiv {x \over \dot x}
&\propto {x_{\rm c}M_{\rm c}\over L_{\star}}
\propto\frac{4-3\mu_c}{\mu_c^{3.27}}.
\end{aligned}
\label{eq:time}
\end{equation}

Although the simple two-layer model gives good results for the convective core mass and radius, it gives a large error, of order $\sim100\%$, for the stellar-radius evolution during the MS. Therefore, we use this simple model to construct a more accurate model of the star that also includes the transition region between the core and the envelope, as presented below.

\section{Three-layer model}
\label{sec:3layers}
We aim to build a model of main-sequence stars composed of three components: a convective, chemically homogeneous core with $\mu_c$, an outer radiative envelope with the ZAMS chemical composition of the star $\mu_e\simeq0.6$, and an intermediate radiative layer where the mean molecular weight smoothly decreases from the core value down to the envelope value. This region extends in mass from the current core mass $M_c$ to the ZAMS core mass, $M_{c,i}\cong M_*/3$. The radius of this region ranges between the radius of the convective core $R_c$ and the inner radius of the unburned envelope $R_e$. 
For this transition zone, we use the relation between mass and $\mu$ from equation~\eqref{eq:mfit} 
\begin{equation}
        \frac{m}{M_c}=( \frac{\mu_c}{\mu})^{5/4}.
    \label{eq:mfitt}
\end{equation}
Similarly, from equation \eqref{eq:kfit} the opacity satisfies 

\begin{equation}
    \frac{\kappa}{\kappa_c}=( \frac{\mu_c}{\mu})^{4/5} (\frac{T}{T_c})^{2/25}.
    \label{eq:kfitt}
\end{equation}

Using equations \eqref{eq:mfitt} and \eqref{eq:kfitt}, and requiring hydrostatic equilibrium and constant luminosity for the transition zone, we require the quantities
\begin{equation}
   \frac{r^2}{Gm\rho}\frac{dP}{dr}\sim \frac{rT}{m^{1/5}},
   \label{eq:hsc}
\end{equation}
and 
\begin{equation}
    L=  \frac{16\pi ac}{3\kappa\rho}T^3 r^2\frac{dT}{dr}\sim \frac{rT^{5}}{Pm^{-1/5}},
    \label{eq:Lc}
\end{equation}
to be constant. 
Because of the weak power-law dependence on mass in both expressions, the temperature and pressure in the transition layer depend mostly on radius. Thus, we can assume for them both a power-law dependence on radius:
\begin{equation}
    T\sim r^{-\alpha}, \quad P\sim r^{-\frac{\alpha}{\nabla_{R_c}}},
    \label{eq:Tpowerlow}
\end{equation}
where $\nabla_{R_c}\equiv\left.\frac{d\ln T}{d\ln P}\right|_{R_c}$ ranges from $2/5$ at the convective region to $1/4$ near the edge of the star, which is described by an $n=3$ polytrope. 
Because of overshooting, there is a small region between the core and the transition region with constant $\mu=\mu_c$ and $\nabla<0.4$, as described in Appendix C. 
Since the $\mu(m)$ profile in the transition region remains fixed during the MS, we expect $\alpha$ and $\nabla_{R_c}$ to also be constants.


Using ideal gas pressure \eqref{eq:pgas}, the density in the transition zone is
\begin{equation}
    \rho=\rho_{ce}(\frac{r}{R_c})^{\frac{\alpha}{\nabla_{R_c}}(\nabla_{R_c}-1)}(\frac{m}{M_c})^{-4/5},
    \label{eq:rhoo}
\end{equation}
where $\rho_{ce}$ is the density at the edge of the core.

Substituting in the continuity equation, we obtain
\begin{equation}
\begin{split}
&\frac{m}{M_c}=\left[c(\frac{r}{r_c})^{3+\frac{\alpha}{\nabla_{R_c}}(\nabla_{R_c}-1)}+(1-c)\right]^{5/9},
\\&c=\frac{5.4b}{3+\frac{\alpha}{\nabla_{R_c}}(\nabla_{R_c}-1)},
 \end{split}
 \label{eq:mmr}
\end{equation}
where $b=\frac{4\pi\rho_{cp}r_c^3}{3M_c}$ is the ratio between the density at the edge of the core and the average core density. 
 Although the quantities \eqref{eq:hsc}, \eqref{eq:Lc}, are not strictly constant under our solution, equations \eqref{eq:mfitt} -- \eqref {eq:kfitt} and \eqref{eq:Tpowerlow}--\eqref{eq:mmr}, we find that they are nearly constant for $\alpha\simeq0.7$,  $\nabla_{R_c}\simeq0.318$, $b\simeq0.8$, $c\simeq2.87$, as shown in Fig.~\ref{fig:HSE}. We also plot equation \eqref{eq:rhoo} using these values and compare it with the MESA simulation in Fig.~\ref{fig:rhocompare}.
 
The conditions at the inner edge of the unburned envelope, where $r=R_e$, are
\begin{equation}
\frac{R_e}{R_c}=0.49\left((\frac{\mu_c}{\mu_e})^{9/4}+1.87\right)^{0.66},
\label{eq:re}
\end{equation}
\begin{equation}
\frac{\rho_e}{\rho_c}=1.63\left((\frac{\mu_c}{\mu_e})^{9/4}+1.87\right)^{-0.98}\frac{\mu_e}{\mu_c},
\label{eq:rho1}
\end{equation}
where for $b=0.8$, $\frac{\rho_{ce}}{\rho_{c}}=0.58$.

\begin{figure}[htbp]
\includegraphics[
    width=\linewidth,
    trim=1cm 0cm 0cm 0cm,
    clip
]{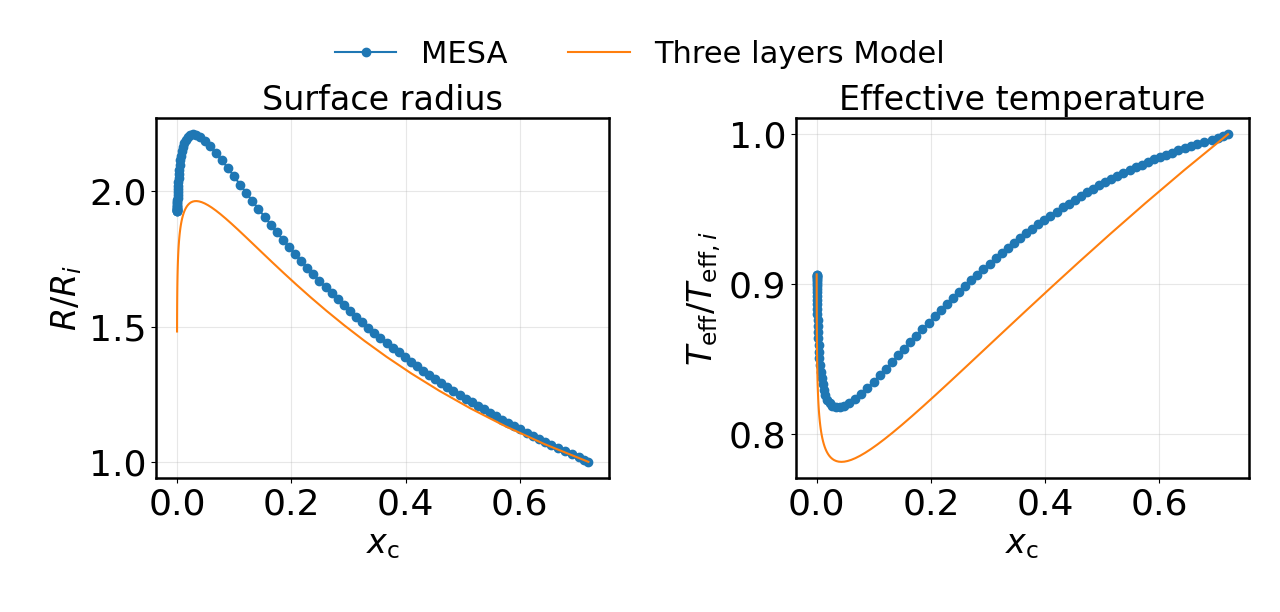}
    \caption{The stellar radius and effective temperature as functions of the core hydrogen fraction from equations \eqref{eq:rstarr}, \eqref{eq:tefff}, compared with the MESA $5M\odot$ run.
Compared with the MESA run, we obtain good agreement, with an error of $\sim10\%$.
    In this three-layer model, the radius has a maximum at $x_{c}$=0.035 and the temperature has a minimum at $x_{c}$=0.045, which resembles the MS "hook".}
    \label{fig:rTeff}
\end{figure}

We now estimate the stellar radius, namely the outer edge of the unburned envelope.  The envelope mass is constant $\simeq 2M_{\star}/3$, and we estimate the stellar radius as 
\begin{equation}
R_{\star}=R_e+\delta{\frac{M_{\star}}{R_e^2\rho_e}}.
\label{eq:RRR}
\end{equation}
We set the coefficient $\delta=1.13$ according to its value at the beginning of the main sequence, where we use the ZAMS conditions \(\frac{R_{\star,i}}{R_{c,i}}\simeq4.5\). In our MESA simulations, for the range $3$--$10\,M_\odot$, we find \(4<\frac{R_{\star,i}}{R_{c,i}}<5\). 
Substituting equations \eqref{eq:re} and \eqref{eq:rho1} we obtain
\begin{equation}
\begin{split}
&\frac{R_{\star}}{R_{\star,i}}=0.175x_c^{1/17}\mu_c^{-17/40}\left[\frac{R_e}{R_c}+3.5(\frac{R_c}{R_e})^{1/2}(\frac{\mu_c}{\mu_e})^{9/4}\right].
\label{eq:rstarr}
\end{split}
\end{equation}
The effective temperature is
\begin{equation}
\begin{split}
   & \frac{T_{\mathrm{eff}}}{T_{\mathrm{eff,i}}}
=2.54x_c^{-1/34}\mu_c^{37/80}\left[\frac{R_e}{R_c}+3.5(\frac{R_c}{R_e})^{1/2}(\frac{\mu_c}{\mu_e})^{9/4}\right]^{-1/2}.
\end{split}
 \label{eq:tefff}
\end{equation}
We plot the stellar radius and effective temperature according to equations~\eqref{eq:rstarr} and \eqref{eq:tefff} in Fig.~\ref{fig:rTeff}, which shows good agreement with results of our MESA simulation.
These expressions demonstrate the behavior of the stellar radius and effective temperature, since they can be written as the product of a function of $\mu_{\mathrm{c}}$ and the weak dependence $x_{\mathrm{c}}^{1/17}$.
Because of this weak dependence, $x_{\mathrm{c}}$ has only a minor effect throughout most of the main sequence, until $\mu_{\mathrm{c}} \rightarrow 4/3$ and $x_{\mathrm{c}} \rightarrow 0$, and therefore an extremum is obtained for some small value of $x_{\mathrm{c}}$.
Using equations \eqref{eq:rstarr} and \eqref{eq:tefff}, we find that $R_{\star}$ reaches a maximum at $x_{\mathrm{c}} \simeq 0.035$, while $T_{\mathrm{eff}}$ reaches a minimum at $x_{\mathrm{c}} \simeq 0.045$, which defines the main-sequence ``hook''.
The ``hook'' occurs before the core is completely exhausted, in the sense that the same physics of a convective core that dominates the burning continues. After the ``hook" the star continues to evolve on the nuclear timescale \eqref{eq:time}, and the ``hook" does not represent a departure from thermal equilibrium, contrary to \cite{renzini1992inflate,Sabhahit2025}. We define the end of the main-sequence as the time when the core becomes less luminous than its surroundings. This occurs only when $x_{\mathrm{c}} \simeq 10^{-4}$, as shown in the next section.
\section{The Transition to Shell Burning}
\label{sec:shell}

\begin{figure}[htbp]
\includegraphics[
    width=\linewidth,
    trim=2.5cm 0cm 0cm 0cm,
    clip
]{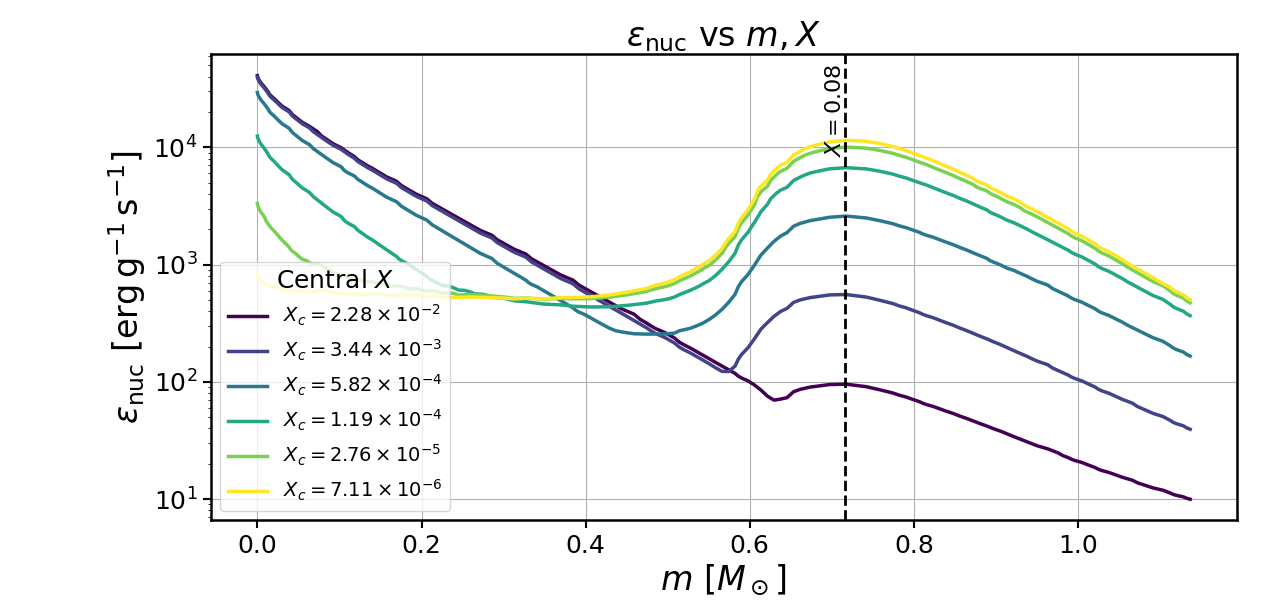}
    \caption{Energy production rate as a function of mass for different central $x_c$ values, from the $M=5\,M_\odot$ MESA run. 
   The figure shows that when $\varepsilon$ decreases significantly in the core, the shell is already well developed, such that there is no isothermal core stage between the MS and the RG phase. The figure also indicates that, during the early stages of shell burning, the shell contains a non-negligible mass.}
    \label{fig:eps}
\end{figure}


At the end of the MS, the hydrogen abundance in the intermediate zone is much higher than in the core, and thus, even though the temperature is lower, this region can produce the same amount of luminosity as the core. We now want to evaluate the luminosity produced in the intermediate zone at that time, when this region becomes a burning shell outside the core, as shown in Fig.~\ref{fig:eps}. 

Thus, we substitute \(\mu_{c}=4/3\), and obtain from equation \eqref{eq:mfit} the core mass and the \(x(m)\) profile outside the core. For simplicity, we use for the shell a polytropic profile (\(n=3\)) with constant $\mu_{c}$, because the burning occurs close to the core where $\mu$ has not decreased significantly.
Using this procedure, we obtain the temperature $T(m)$ and density $\rho(m)$ profiles, as well as the hydrogen fraction profile \(x(m)\) of the shell. We use the ratio between the conditions at the edge of the core and the central values $\frac{\rho_{ce}}{\rho_{c}}=0.58,\frac{T_{ce}}{T_{c}}=0.69$ derived in the previous section.
Using these assumptions, the nuclear luminosity produced in the shell is
\begin{equation}
\begin{split}
L_{\rm shell}
&\simeq
\epsilon_1 z
\int_{M_{\rm core}|_{\mu=4/3}}^{M_{\star}/3}
x(m)\rho(m)T(m)^{\nu}\,dm
\\
&\simeq
2.5\times10^{-5}\,\epsilon_1 z\,
\rho_c^2 T_c^{\nu} r_{\rm c}^3.
\end{split}
\end{equation}

Because of the fixed shape of the shell, the factor $2.5\times10^{-5}$ does not depend on the central conditions of the star and can therefore be computed once.
The luminosity generated in the adiabatic core is
\begin{equation}
    L_{\mathrm{core}}\simeq 0.10 \epsilon_1 z x_c \rho_c ^2T_c^{\nu}r_{c}^3.
\end{equation}
The ratio between the two luminosities during the later stages of the MS is therefore
\begin{equation}
    \frac{L_{core}}{L_{shell}}\simeq 4000x_{c}, 
\end{equation}
and the transition to shell burning occurs when the luminosities are equal, at
\begin{equation}
    x_{c,end}\simeq2.5\times10^{-4}.
    \label{eq:xend}
\end{equation}

In our MESA runs for masses in the range \(3\text{--}10\,M_\odot\) we obtain similar values \(x_{c,end}= 1\text{--}3.4\times10^{-4}\).
Using equation \eqref{eq:time} we can compute the fraction of the time the star spends between the ``hook'' and the end of the MS
\begin{equation}
    \frac{t_{\mathrm{after\ hook}}}{t_{\mathrm{until\  hook}}}=\frac{\int_{x_c=0.045}^{x_c=2.5\times10^{-4}}\mu_c^{-2.27}dx}{\int_{x_c=0.75}^{x_c=0.045}\mu_c^{-2.27}dx}\simeq 0.020,
\end{equation}
independent of the stellar mass. In comparison, in our MESA runs for masses in the range \(3\text{--}10\,M_\odot\) this ratio is between $0.010$ to $0.018$.

Using the $T(m)$ and $x(m)$ profiles outside the core, we also find the values at the shell peak
\begin{equation}
    x_{peak}\sim0.05\quad m_{peak}\simeq1.2M_{c}.
\end{equation}
This shows that the burning shell is not very thin in its early stages.

An important conclusion arises from our description of the end of the MS: the transition from a burning core to a burning shell is continuous, so that there is no stage in which the core is isothermal and the shell has not yet ignited, as suggested by the assumptions underlying the Schönberg–Chandrasekhar mass limit \citep{Schonberg1942}.
We also show that the shell mass is comparable to the core mass, and therefore the shell cannot be neglected in modeling the star at this stage, contrary to what is suggested by the Schönberg–Chandrasekhar calculations. Moreover, from equation (\ref{eq:Mc_final_frac}) we see that the core mass at the end of the main sequence is already equal to or larger than the Schönberg–Chandrasekhar mass. Therefore, a model in which 
an isothermal core grows until it reaches the S–C mass as described by \citet{Kippenhahn2012} is not relevant for describing the end of the main sequence.

\section{Discussion}
\label{sec:dis}
For stars more massive than $5\,M_\odot$, we obtain core masses larger than expected from equation~\eqref{eq:Mc_from_Mtot_simple}. This is because we neglect radiation pressure, which becomes increasingly important as the stellar mass grows. 
In our mass range, at the end of the main sequence,
\(
\frac{P_{rad}}{P_{gas}}
\)   varies between 0.02 and 0.18 for masses between $3$ and $10\,M_\odot$. 
When radiation pressure becomes important, the effective $\gamma$ in the core approaches $4/3$ instead of $5/3$. 
This value of $\gamma$ reduces $\frac{dT}{dr}$ and therefore increases the luminosity produced in the core. As a result, the core mass increases beyond the value predicted by our assumptions.
Moreover, the varying importance of radiation pressure modifies the initial mass ratio
\( \frac{M_{e,i}}{M_{c,i}},
\) which varies between 1.6 and 2.5, so that the $\mu(m)$ profile derived from equation (\ref{eq:Mc_from_Mtot_simple}) becomes slightly mass-dependent. 
For the same reasons, $x_{\mathrm{c,end}}$ also varies slightly over this mass range.

Another important factor affecting our results is the treatment of overshooting.
Although we do not derive a detailed treatment of overshooting here, we can still predict its influence on the stellar properties.
We show that overshooting creates a small layer between the convective region and the transition zone where $\mu$ begins to decrease. In that region, $\nabla$ decreases, which affects the power laws of temperature and pressure in the transition zone. Overshooting also increases the core mass and radius, which is expected to increase both the luminosity and the stellar radius during the main sequence, while also shortening the evolutionary timescale at this stage, as shown in equations~(\ref{eq:Lfinal}), (\ref{eq:rstarr}), and (\ref{eq:time}). This is consistent with previous studies \citep{Kippenhahn2012,Anders2023}.  


The metallicity $Z$ enters our model mainly through the nuclear energy-production rate. 
Nevertheless, we expect changes in metallicity to have only a weak effect on our main results, in particular on the core-mass relation in equation~\eqref{eq:Mc_from_Mtot_simple} and on the value of $x_{\rm c,end}$ in equation \eqref{eq:xend}. 
We validate this expectation using MESA simulations of a $5\,M_{\odot}$ star over the metallicity range $Z=0.0002$--$0.02$.

The main differences among the metallicity models appear in the ZAMS values of the effective temperature and luminosity. 
These differences change the time the star spends on the main sequence by about 20\%.
 Metallicity also affects the opacity near the stellar surface and therefore changes the evolution of the stellar radius and effective temperature during the main sequence by about 10\% over the metallicity range explored in this work.
 
\section{Summary}
\label{sec:sum}
We derive a simple analytical description of the evolution of $3\text{--}10\,M_\odot$ main-sequence stars.
Our model consists of three layers: (i) a convective (adiabatic) core, with mean molecular weight $\mu=\mu_{c}$, and a known nuclear burning rate $\epsilon(\rho,T)$, dominated by Thomson opacity with a small temperature dependence due to metals. The core terminates where the total nuclear energy production equals the radiative luminosity. (ii) A radiative intermediate region with power-law profiles of $\mu(m)$,  $T(r)$ and $P(r)$, and constant luminosity, which extends from the core until $\mu=\mu_e$. (iii) A radiative envelope with constant luminosity $L_{\star}$ and mean molecular weight $\mu_{e}=0.6$. 
Finally, we require the total mass to be $M_\star$.

Using this model we derive expressions for the convective core mass~\eqref{eq:Mc_from_Mtot_simple} and radius~\eqref{eq:rrr}, as a function of $\mu_c$.
For the stellar luminosity we obtain $  L_{\star}\propto \mu_{c}$, and we determine the evolutionary timescale at any stage along the main sequence \eqref{eq:time}.
We also derive analytical expressions for the stellar radius \eqref{eq:rstarr} and effective temperature \eqref{eq:tefff}.
We show that the main-sequence ``hook'', defined as the minimum of the effective temperature, does not correspond to a complete exhaustion of the hydrogen in the core and occurs once $x_{\mathrm{c}} \simeq 0.045$. 
The same physics of hydrogen burning in a convective core described by equations \eqref{eq:Mc_from_Mtot_simple}, \eqref{eq:ttt}--\eqref{eq:time} and \eqref{eq:rstarr}--\eqref{eq:tefff}, continues to govern the evolution after the hook, until much smaller values of $x_{\mathrm{c,end}}=2.5\times 10^{-4}$, at which point the region surrounding the core becomes as luminous as the core itself. The "hook" is simply a manifestation of the minimum of equation (\ref{eq:tefff})  as a function of $x_{\rm c}$. We find the terminal MS core mass \(
M_{\mathrm{c,end}} \simeq 0.11\,M_{\mathrm{\star}}.
\)
This value is numerically close, but unrelated to the Schönberg--Chandrasekhar mass. 
During the late stages of evolution, the core remains convective, rather than isothermal, and the Schönberg--Chandrasekhar limit is not relevant to shell ignition. 

This value of the core mass at the end of the MS can be useful, for example, for estimating the mass of helium white dwarfs formed in close binary systems shortly after the main sequence \citep{Tauris2000,Tauris2011,ShaoLi2012,Misra2020,BareliGinzburg2026}.

The expression for the core mass~\eqref{eq:Mc_from_Mtot_simple}, also provides the $x(m)$ profile outside the core, which is generated by the shrinking of the core mass. This profile is important for modeling both the transition region between the core and the envelope, and the burning shell that develops in this region at the end of the MS. This profile allows us to identify the value of $x_{\mathrm{c,end}}$ at which the shell produces the same luminosity as the core.
The derivation of the $x(m)$ profile outside the core can also be useful for modeling the burning shell in the transition stage between the main sequence and the red-giant phase, namely the Hertzsprung gap, by determining the width of the shell during that stage. It also influences the red giant phase, because the burning shell is located at a smaller mass coordinate than the ZAMS convective core mass, even at advanced stages of the red giant phase.

\section*{Acknowledgments}
This research was partially supported by an NSF-BSF grant and a GIF grant. We thank Ehud Nakar for useful discussions.
\clearpage
\raggedbottom
\appendix
\section{Core structure in dimensionless form}
In this appendix, we solve the relevant equations for the two-layer model, including factors of order unity. Assuming central conditions \(T_c\), \(\rho_c\), and \(\mu_c\), we can find the properties of the convective core using:

Hydrostatic equilibrium:
\begin{equation}
\frac{dP}{dr} = -\frac{G m \rho}{r^2}
\label{eq:hs}
\end{equation}

Conservation of mass:
\begin{equation}
\frac{dm}{dr} = 4\pi \rho r^2
\label{eq:mass}
\end{equation}

and the adiabatic relation for an ideal gas:
\begin{equation}
P = P_c \left(\frac{\rho}{\rho_c}\right)^{5/3}.
\label{eq:poly}
\end{equation}

From these equations we obtain the temperature and density profiles of the core \(T(r)\), $\rho(r)$.  
From these profiles we can compute the nuclear and the radiative luminosity using equation \eqref{eq:ths}. The point where the two luminosities are equal defines the edge of the core.

For the radiative envelope, we replace equation~\eqref{eq:poly} with equation~\eqref{eq:ths}, assuming constant luminosity $L_{\star}$.

For any $\rho_c,  T_c$ we can define two characteristic radii, one from   \ref{eq:2},\begin{equation}
r_{c1} = \sqrt{\frac{k_B}{ G \mu_{c} m_H\rho_c}T_c},
\label{eq:rhs}
\end{equation}
and one from equation \ref{eq:ths}:
\begin{equation}
r_{c2}=\sqrt{\frac{ac T_c^4 }{\kappa_c\rho_c^2 \epsilon}}
\end{equation}
Demanding both HS and thermal equilibrium is equivalent to requiring $r_{c1}\sim r_{c2}\sim r_c$.  
Equating the two radii gives the characteristic temperature:

\begin{equation}
T_c=( \frac{k_B \kappa_c \rho_c^2 x_{c} z\epsilon_1}{ G \mu_{c}m_H ac})^{\frac{1}{3-\nu}}
\end{equation}
Using $T_c, r_{c}$, \(m_{c}=\rho_c r_{c}^3\), and \(L_{\star}=\rho_c^2 T_c^\nu r_{c}^3\),  assuming constant opacity and a chemically homogeneous star, we can write \eqref{eq:hs}, \eqref{eq:mass}, \eqref{eq:poly}, and \eqref{eq:ths} in dimensionless form.  For   
\(\bar T=T/T_c, \bar r=r/r_{c}, \bar \rho=\rho/\rho_c, \bar{m}=m/m_{c}, \bar L =L/L_{\star} \), we get:

\begin{equation}
\frac{d\bar T \bar \rho }{d\bar r}=-\frac{\bar m\bar \rho }{\bar r^2}
\label{eq:hsles}
\end{equation}

\begin{equation}
\frac{d\bar m}{d\bar r}=4\pi \bar\rho\bar{r}^2
\label{eq:dmles}
\end{equation}

\begin{equation}
\frac{\bar T}{\bar T_c}=\bar \rho^{2/3}
\label{eq:polyles}
\end{equation}

\begin{equation}
\bar L_{nuc}
=\int\bar \rho ^2\bar T^\nu \bar r^2 d\bar r
\label{eq:Lnucles}
\end{equation}

\begin{equation}
-  \frac{\bar r^2}{3\bar\rho} \frac{d\bar T^4}{d\bar r}=\bar L_{rad}
\label{eq:Lradles}
\end{equation}

To solve these equations, we proceed as follows: We choose an arbitrary $\rho_c$ and define $\bar\rho_c=1$, then solve the dimensionless equations for different values of $\bar T_c$. For each $\bar T_c$, we solve equations~\eqref{eq:hsles}, \eqref{eq:dmles}, and \eqref{eq:polyles}, and compute $\bar L_{rad}$ and $\bar L_{nuc}$ to find the edge of the core at $\bar r_{c}$ where \(\bar L_{nuc}=\bar L_{rad}\).  
From this point onward, we replace equation~\eqref{eq:polyles} with equation~\eqref{eq:Lradles}.

Eventually, we obtain a particular $\bar T_c$ such that simulations with larger $\bar T_c$ end too early and reach infinite density at their endpoint, whereas solutions with smaller $\bar T_c$ become isothermal and do not terminate anywhere.  
These two situations indicate that the star is not in thermal equilibrium. In the first case \(L_{nuc}>L_{rad}\) and the star will expand and cool until $\bar T_c$ and $L_{nuc}$ decrease, and vice versa.  
The value of $\bar T_c$ between these two cases is the one that satisfies both hydrostatic and thermal equilibrium, and we find $\bar T_c=1.37$.

Equations~\eqref{eq:hsles}--\eqref{eq:Lradles} are independent of the chemical composition, the opacity value, and the initial central conditions.  

\subsection{Non-homogeneous stars with chemically evolved cores}
We can repeat the same process for non-homogeneous stars.
We can model these stars as a convective core with $\mu_{c}$ and a radiative envelope with $\mu_{e}$, assuming an unrealistic jump in $\mu$.  
This model differs from the homogeneous one in the density jump which results from equation \eqref{eq:3}, and in equation \ref{eq:hsles}, which becomes:

\begin{equation}
\frac{d\bar T \bar \rho }{d\bar r}=-\frac{\bar m\bar \rho }{\bar r^2}\frac{\mu_{e}}{\mu_{c}}
\end{equation}

where \(1<\frac{\mu_{c}}{\mu_{e}}<\frac{\mu_{He}}{\mu_{H}}\simeq2.2\).

If we also account for the jump in opacity implied by Thomson opacity, then \ref{eq:Lradles} becomes:

\begin{equation}
-  \frac{\bar r^2}{3\bar\rho} \frac{d\bar T^4}{d\bar r}
=\bar L\frac{\kappa_{e}}{\kappa_{c}}
\end{equation}

where \(1<\frac{\kappa_{e}}{\kappa_{c}}< 1.7\).

We find that for $\mu_{c}\simeq\mu_{He}$ we obtain $\bar T_c=1.36$, such that within the MS core chemical composition range, $\bar T_c$ varies by less than 1\%. Therefore, we can write the central temperature:

\begin{equation}
T_c = 
1.37\left(\frac{k_B z \epsilon_1}{Gacm_H}(1+\frac{2}{\mu_c})x_{c}\rho_c^{2} \mu_{c}^{-1}\right)^{\frac{1}{3-\nu}}. 
\label{eq:T00}
\end{equation}

Substituting equation~\eqref{eq:polyles} into equations~\eqref{eq:hsles} and \eqref{eq:Lradles} gives:

\begin{equation}
\frac{d\bar  T}{d\bar r}=-\frac{2\bar m}{5\bar r^2}
\rightarrow 
\bar L_{rad}=  \frac{4\bar T^3\bar m }{3\bar\rho} 
\sim \bar T^{3/2} \bar r^3
\label{eq:dtles}
\end{equation}

From \ref{eq:dtles}, considering only the convective regime, we can define the shapes of the two luminosities.  
Near the center of the core, where \(\bar T\simeq\bar T_c\), we get \(L_{rad}\sim\bar r^3\), the same as $L_{nuc}$, as can be seen from \ref{eq:Lnucles}.  
Farther away, while $L_{nuc}$ becomes constant, $L_{rad}$ reaches a maximum and then declines until it reaches zero, as does $\bar T$.

If there is no intersection point between the two luminosities before $L_{rad}$ reaches its maximum, there is no transition to a radiative regime, and the star will be unable to transfer its energy. This causes the star to expand and cool, decreasing $L_{nuc}$ until it reaches the maximum of $L_{rad}$.

Because of these fixed shapes, the only quantity that determines $\bar r_{c}$ is the ratio $\eta$ between the two luminosities at the center:
\begin{equation} 
\eta=\frac{L_{nuc}}{L_{rad}}|_{r\rightarrow0}=\frac{\gamma}{\gamma-1}\frac{3k_B}{16\pi aG\mu c}\frac{\epsilon\kappa\rho}{T^3}= \bar T_c^{\nu-3}\frac{15}{32\pi}\simeq4.39
\end{equation}

Thus, we can treat $\eta$, and therefore $\bar r_{\rm c}\simeq0.55$ and $\bar m_{\rm c}\simeq0.56$, as constants during the MS and thus:

\begin{equation}
M_{c}=0.56\rho_c r_{c} ^3
    \label{eq:Mcored0}
\end{equation}

From equations~\eqref{eq:mfit}, \eqref{eq:kfit}, \eqref{eq:T00}, and \eqref{eq:Mcored0} we get the relation between $\rho_c$ and $\mu_{c}$:
\begin{equation}
\begin{split}
\rho_c^{-1/2} T_c^{3/2}\mu_{\rm c}^{-3/2}
\sim\mu_c^{-7/5}
\rightarrow
\rho_c\sim \mu_c^{1/40}x_c^{-3/17}
\end{split}
\label{eq:rho}
\end{equation}
Similarly the central temperature is
\begin{equation}
T\sim (\mu_c^{14/85}x_c^{-1/17}).
\end{equation}
Substituting equation~\eqref{eq:rho} into equation~\eqref{eq:Mcored0} gives
\begin{equation}
\begin{split}
R_{\rm c}
&\sim \left(\frac{M_{\rm c}}{\rho_c}\right)^{1/3}
\sim \mu_c^{-17/40}x_c^{1/17}.
\end{split}
\label{eq:rcore}
\end{equation}

\section{Average $\mu$}
From equation~\eqref{eq:Mc_from_Mtot_simple}, we have for the region outside the core
\begin{equation}
m(\mu)=\frac{M_{\mathrm{\star}}}{1+C\mu^2},
\qquad
C=\frac{2.1}{\mu_{e}^2},
\end{equation}
where $\mu_{\rm e}=0.6$.
Differentiating with respect to $\mu$ gives
\begin{equation}
dm 
= -M_{\mathrm{\star}}\frac{2C\mu}{(1+C\mu^2)^2}\,d\mu.
\end{equation}

The mass-weighted average of $\mu$ in the intermediate zone is therefore
\begin{equation}
\langle\mu\rangle_p
=
\frac{\int_{\mu_e}^{\mu_c} \mu\,dm}{\int_{\mu_e}^{\mu_c} dm}
=
\frac{\int_{\mu_e}^{\mu_c}
\frac{-2C\mu^2}{(1+C\mu^2)^2}\,d\mu}
{\left[\frac{1}{1+C\mu^2}\right]_{\mu_e}^{\mu_c}}.
\end{equation}

The numerator can be integrated analytically:
\begin{equation}
\int \frac{-2C\mu^2}{(1+C\mu^2)^2}d\mu
=
-\frac{1}{\sqrt C}\tan^{-1}(\sqrt C\,\mu)
+
\frac{\mu}{1+C\mu^2}.
\end{equation}

Substituting, we obtain the final expression:
\begin{equation}
\langle\mu\rangle_p
=
\frac{
\left[
-\frac{1}{\sqrt C}\tan^{-1}(\sqrt C\,\mu)
+
\frac{\mu}{1+C\mu^2}
\right]_{\mu_e}^{\mu_c}
}{
\left[
\frac{1}{1+C\mu^2}
\right]_{\mu_e}^{\mu_c}
}.
\end{equation}

We can now treat the envelope as a region with constant average $\mu$:
\begin{equation}
    \langle\mu\rangle_{e}= \frac{\langle\mu\rangle_p(M_{c,i}-M_{c})+0.6(M_{\star}-M_{c,i})}{M_{\star}-M_{c}}
\end{equation}

\section{The overshooting layer}
Because of the overshooting, between the edge of the convective region and the transition zone where $\mu$ begins to decrease, there is a radiative layer with constant $\mu=\mu_c$. In this layer, by definition 

\begin{equation}
    \nabla\sim\frac{\rho}{ mT^3}<0.4,
\end{equation}
where we use HSE and constant luminosity \eqref{eq:ths}.
Substituting $\rho\sim m/r^3$ gives for that region \begin{equation}
    T\sim m^{1/4}/r, \nabla\sim  1/m^{3/4}.
    \label{eq:nabla}
\end{equation}
For the exponential-overshooting prescription used in our MESA simulations, the diffusion coefficient outside the convective region is
\begin{equation}
    D=D_0\exp\left[-\frac{a(r-R_{conv})}{H_{p,conv}}\right],
\end{equation}
where $D_0=v_{conv}H_p/3$ is the diffusion coefficient at a point just inside the edge of the convective region, at $r_0=R_{conv}-0.004H_{p,conv}$, and $a=142.8$ is a constant. The subscript $conv$ denotes the edge of the convective region. From equation \eqref{eq:Tpowerlow} $R_{conv}/H_{p,R_{conv}}=\nabla_{R_c}/\alpha$. Equating the nuclear time-scale \eqref{eq:time} and the diffusion time at the edge of the core, the ratio between the radii during the MS is 
\begin{equation}
    \frac{R_{c}}{R_{conv}}=1+\frac{\nabla_{R_c}}{a\alpha}\ln\left(\frac{D_0t(x)}{R_c^2}\right)=const,
\end{equation}
where we use the fact that $a$ is large, and thus the logarithmic term has a minor importance. We also assume that $\nabla_{R_C}$ and $\alpha$ are constants during the MS and therefore, for this particular overshooting prescription, we find that the overshooting-layer radius and mass are constant fractions of the core radius and mass. Substituting this result into equation~\eqref{eq:nabla} justifies the
assumption that $\nabla_{R_c}$ and $\alpha$ remain constant.
\makeatletter
\setlength{\@fptop}{0pt}
\setlength{\@fpsep}{10pt}
\setlength{\@fpbot}{0pt plus 1fil}
\makeatother

\begin{figure}
    \centering
    \includegraphics[width=1\linewidth, trim=1.5cm 0cm 0cm 0cm,
    clip]{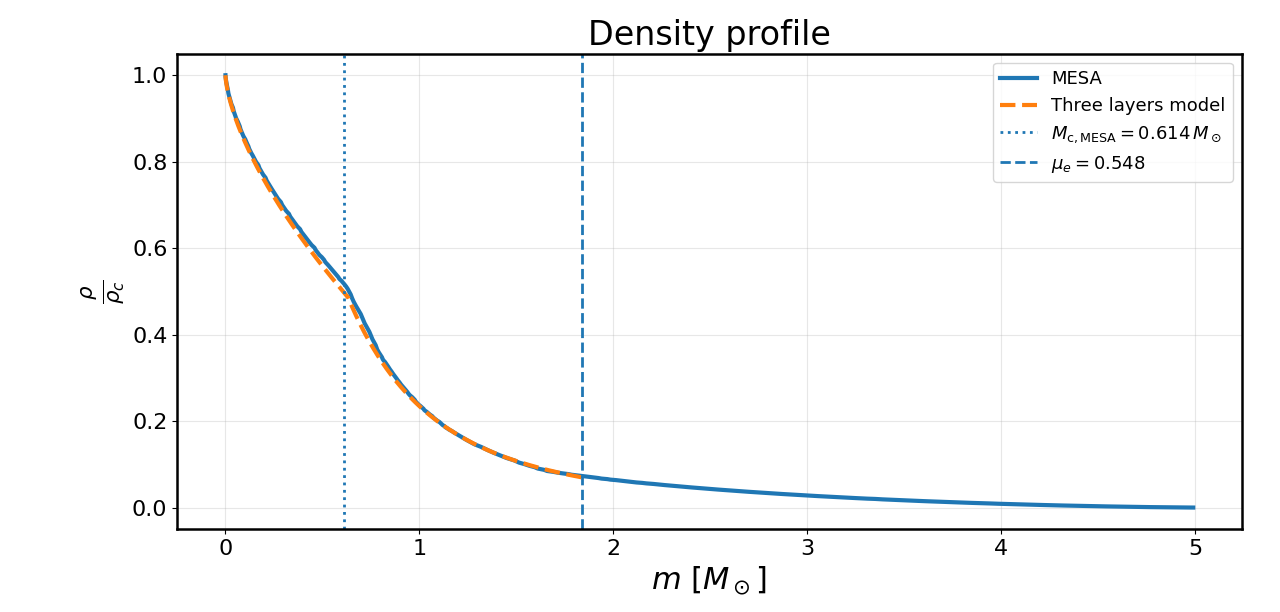}
    \caption{Density divided by the central density as a function of mass coordinate for $M=5M_{\odot}$ MESA simulation. In orange, we plot the result from our three-layer model, starting with a convective core that terminates at the point where $b=\frac{4\pi \rho_{cp}r_c^3}{3M_c}=0.801$  and continues with equation \eqref{eq:rho1}  until $\mu=0.6$. The vertical lines bound the transition region where $\mu$ changes with $m$.}
    \label{fig:rhocompare}
\end{figure}

\begin{figure}
    \centering
    \includegraphics[width=1\linewidth, trim=2.5cm 0cm 0cm 0cm,
    clip]{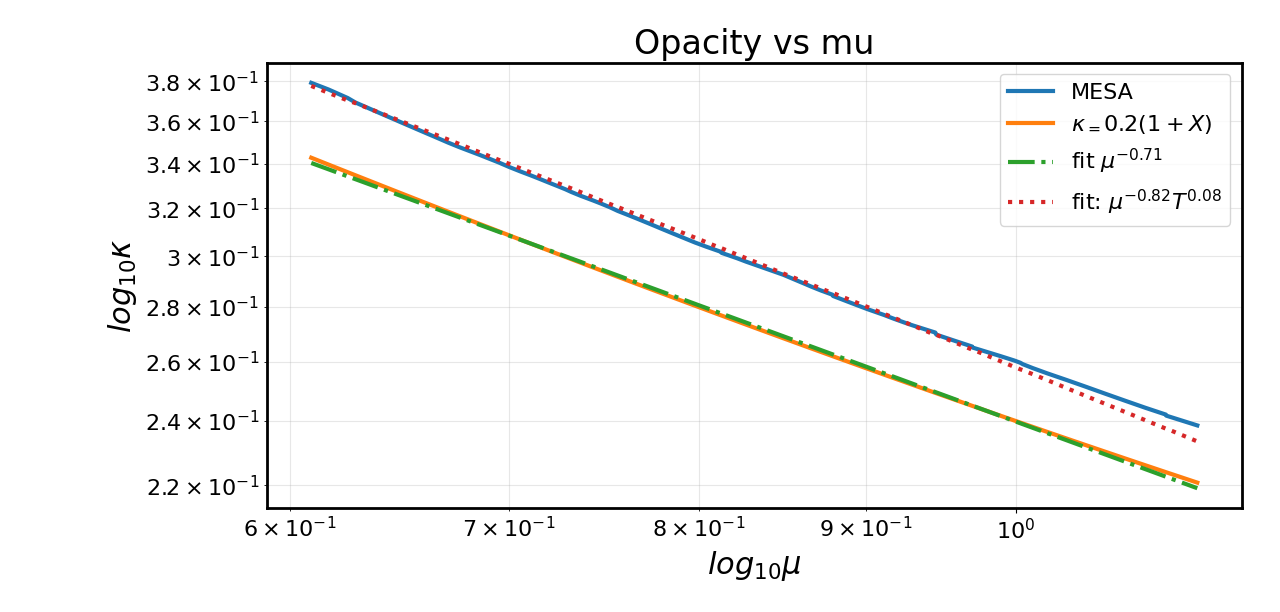}
    \caption{Opacity from the $M=5M_{\odot}$ MESA simulation as a function of $\mu_c$ for the MS $\mu_c$ range $0.6<\mu_c<4/3$, in log scale. In orange we plot Thomson opacity in this range.
    We plot power-law fits for both curves $\kappa_{\rm MESA}\sim\mu^{-0.82}T^{0.08}$,$\kappa_{\rm Th}\sim\mu^{-0.71}$, which show that the opacity in this star is mainly the Thomson opacity.}
    \label{fig:kappa}
\end{figure}

\begin{figure}
    \centering
    \includegraphics[width=1\linewidth, trim=1cm 0cm 0cm 0cm,
    clip]{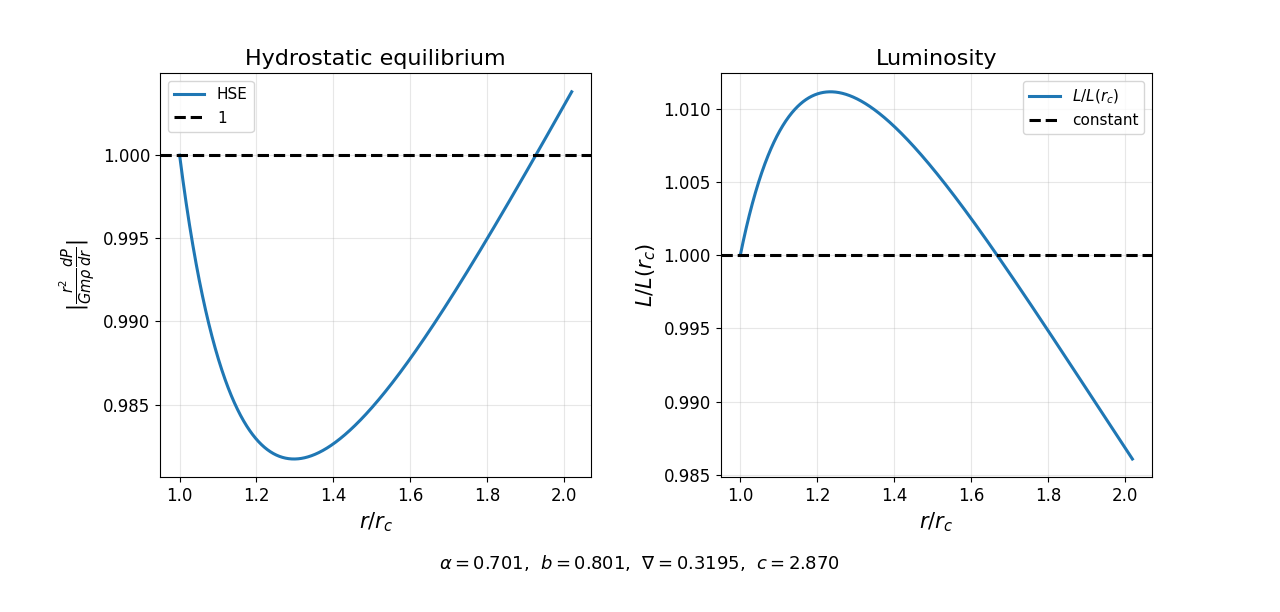}
    \caption{HSE and luminosity at the transition region with $\alpha=0.701, \nabla_{R_c}=0.319$. $L$ varies by $2.5\%$ in that region, while the HSE quantity varies by about $2\%$.}
    \label{fig:HSE}
\end{figure}
\clearpage

\bibliographystyle{aasjournal}
\bibliography{main2}
\end{document}